\documentclass[conference,a4paper,twocolumn]{IEEEtran}

\IEEEoverridecommandlockouts

\ifCLASSINFOpdf
\else
\fi
\usepackage[cmex10]{amsmath}
\usepackage[ruled,vlined]{algorithm2e}  

\usepackage{subfigure}

\usepackage{stfloats}
\usepackage{url}

\usepackage{graphicx}
\usepackage{amssymb}
\usepackage{hyperref}
\usepackage{xcolor}
\usepackage[T1]{fontenc}
\usepackage[utf8]{inputenc}
\usepackage{breakurl}
	
	\usepackage{pgfplots}
  \pgfplotsset{compat=newest}
\newlength\figurewidth

\def\RR{\mathbb{R}}

\def\CC{\mathbb{C}}

\def\x{\vect{x}}
\def\u{\vect{u}}
\def\vv{\vect{v}}
\def\y{\vect{y}}
\def\z{\vect{z}}
\def\w{\vect{w}}

\def\c{\vect{c}}

\def\t{\vect{t}}
\def\Mr{M_\text{R}}
\def\Mh{M_\text{H}}
\def\Ml{M_\text{L}}

\newcommand{\norm}[1]{\|#1\|}

\newcommand{\vect}[1]{\mathbf{#1}} 

\newcommand{\argmin}{\mathop{\operatorname{arg~min}}}

\newcommand{\soft}{\mathrm{soft}}
\newcommand{\proj}{\mathrm{proj}}

\newcommand{\sgn}{\mathrm{sgn}}

\newcommand{\dsdr}{\ensuremath{\Delta\text{SDR}}}
\newcommand{\tc}{\ensuremath{\theta_\text{c}}}
\newcommand{\wa}[1]{\ensuremath{\w_{\text{ATH#1}}}}

\newcommand{\qm}[1]{``#1''}  

\begin{document}



\title{Psychoacoustically Motivated Audio Declipping \\Based on Weighted $\ell_1$ Minimization}


\author{\IEEEauthorblockN{
Pavel Z\'{a}vi\v{s}ka, Pavel Rajmic, Ji\v{r}\'{i} Schimmel
}
\IEEEauthorblockA{
Signal Processing Laboratory\\
Brno University of Technology, 
Czech Republic\\
Email: xzavis01@vutbr.cz, rajmic@vutbr.cz, schimmel@feec.vutbr.cz
}
\thanks{The authors thank O.\ Mokr\'{y} for valuable consultations and proofreading. The work was supported by the joint project of the FWF and the Czech Science Foundation (GA\v{C}R): numbers I3067-N30 and 17-33798L, respectively. Research described in this paper was financed by the National Sustainability Program under grant LO1401. Infrastructure of the SIX Center was used.
}}


%


\maketitle


\begin{abstract}
A novel method for audio declipping based on sparsity is presented.
The method incorporates psychoacoustic information by weighting the transform coefficients in the $\ell_1$ minimization.
Weighting leads to an improved quality of restoration while retaining a low complexity of the algorithm. 
Three possible constructions of the weights are proposed, based on the absolute threshold of hearing, 
the global masking threshold and on a~quadratic curve.
Experiments compare the restoration quality according to the signal-to-distortion ratio (SDR) and PEMO-Q objective difference grade (ODG)
and indicate that with correctly chosen weights, 
the presented method is able to compete, or even outperform, the current state of the art.
\end{abstract}


\begin{IEEEkeywords} 
Declipping; Psychoacoustics; Restoration; Sparsity
\end{IEEEkeywords}

%
\IEEEpeerreviewmaketitle

\vspace{-0.4em}
\section{Introduction}
The so-called clipping is a non-linear form of signal distortion usually appearing when the signal exceeds its allowed dynamic range.
This unwanted phenomenon is most commonly present in audio signals, but clipping can affect all kinds of signal. 
Clipping has a great negative effect on the perceptual quality of audio signal \cite{Tan2003}, 
it reduces the accuracy of automatic voice recognition \cite{Malek2013:Blind.compensation,Tachioka2014:Speech.recog.performance} 
or it may even potentially damage the audio speaker \cite{Patent_clipping_amplifier}.
%
To enhance the perceived audio quality and to reduce the negative effects of clipping, it is necessary to perform the recovery of clipped samples.
This process is usually termed \emph{declipping}.

In the past, several declipping methods using various approaches were introduced.
The first attempts were based on autoregressive modeling \cite{javevr86} or on the knowledge of the original signal bandwidth \cite{abel91_declipping}.
Statistical approaches were exploited in \cite{Godsill2001:StatisticalAudioRestoration,FongGodsill2001:MonteCarlo} and more recently in \cite{Chantas2018:Inpainting_Variational_Bayesian_Inference}.
Also, a method based on the Hankel matrix rank minimization was used in \cite{Takahashi2013:Hankel_matrix_declipping}.

Well-performing methods for the tasks of audio restoration proved to be methods using the sparse prior; 
for instance methods based on Orthogonal Matching Pursuit (OMP) \cite{Adler2011:Declipping},
convex optimization \cite{Weinstein2011:DeclippingSparseland,Defraene2013:Declipping.perceptual.compressed.sensing},
social sparsity \cite{SiedenburgKowalskiDoerfler2014:Audio.declip.social.sparsity},
 or non-convex methods based on the Alternating Directions Method of Multipliers (ADMM) \cite{Kitic2013:Consistent.iter.hard.thresholding,KiticBertinGribonval2014:AudioDeclippingCosparseHardThresholding,Kitic2015:Sparsity.cosparsity.declipping,ZaviskaRajmicPrusaVesely2018:RevisitingSSPADE,ZaviskaRajmicMokryPrusa2019:SSPADE_ICASSP}.
Besides sparsity-based approaches, methods using non-negative matrix
factorization were also adapted for audio declipping \cite{BilenOzerovPerez2015:declipping.via.NMF, Bilen2018:NTF_audio_inverse_problems}.
To the best of our knowledge, deep learning techniques have only been used for image declipping \cite{HonigWerman2018:Image.decliping}, not for audio.


All the methods mentioned above (except for \cite{Defraene2013:Declipping.perceptual.compressed.sensing}) have not exploited any additional information in order to achieve the best possible \emph{perceived} audio quality of the restored signal, which does not necessarily fully coincide with the physical signal restoration quality.
In \cite{Defraene2013:Declipping.perceptual.compressed.sensing}, the authors utilized the effect of simultaneous masking and used the MPEG psychoacoustic model to weight time-frequency coefficients during the restoration process. 
Such an approach discourages the introduction of distinctively audible signal components
(where the masking threshold is low),
which are not likely to be present in the original signal, and signal components that are less audible
(masking threshold is high)
are tolerated to a~greater extent.
Unfortunately, the algorithm presented in \cite{Defraene2013:Declipping.perceptual.compressed.sensing} is not \emph{fully consistent} 
which means that it allows deviating the reconstructed samples from the reliable samples
(i.e., samples that have not been clipped).

In this paper, a fully consistent declipping method based on weighted $\ell_1$~minimization is presented.
First, the declipping problem is formulated and the minimization task is constructed in Sec.~\ref{sec:problem_formulation}.
The problem is then numerically solved by the Douglas-Rachford algorithm \cite{combettes2007douglas}, described in Sec.~\ref{sec:alg_solution}. 
Three different approaches incorporating psychoacoustical information in 
the audio declipping are introduced in Sec.~\ref{sec:weights}.
Finally, Sec.~\ref{sec:experiments} reports on the experiments that have been run and the results obtained.



\newcommand{\ana}{A}
\newcommand{\syn}{D}


\section{Problem formulation}
\label{sec:problem_formulation}
In the case of the so-called hard clipping,
which is the subject of interest in this paper,
the signal exceeding the prescribed dynamic range $[-\tc, \tc]$ is limited in amplitude such that
\begin{equation}
y_n = \left\{
\begin{aligned}
&x_n &\text{for} \hspace{1em} &|x_n| < \tc, \\
&\tc \cdot \sgn(x_n) &\text{for} \hspace{1em}  &|x_n| \geq \tc,
\end{aligned}
\right.
\label{eq:clipping}
\end{equation}
where $\x \in \RR^N$ denotes the original (clean) signal and $\y \in \RR^N$ the observed clipped signal.
The limiting constant $\tc$ is referred to as the clipping threshold.
The subscript $n$ in $y_n$ refers to the $n$-th sample of signal $\y$, the same as $(\y)_n$ does.

According to \eqref{eq:clipping}, it is possible to divide the signal samples into three disjoint sets
$R,H,L$ such that $R\cup H\cup L = \{1,\ldots,N\}$ and, correspondingly,
to distinguish the \emph{reliable} samples (not influenced by clipping), samples \emph{clipped from above} to the high clipping threshold $\tc$ and samples \emph{clipped from below} to the low clipping threshold ($-\tc$), respectively.
To select only samples from one specific set, the respective restriction operators $\Mr$, $\Mh$ and $\Ml$ are used.

In the restoration task, it is natural to desire that the recovered signal
should not differ from the clipped signal $\y$ at the reliable positions, 
and at the clipped positions its samples should lie above $\tc$ or below $-\tc$.
Such a task is ill-posed, since the solution is not unique. 
Therefore, considering some additional information about the signal is crucial. 
Based on the fact that most of the musical signals are approximately sparse with respect to a certain time-frequency transform, 
the restoration task can be formulated as finding a signal
whose coefficients
will be of the highest sparsity.
As the true sparsity measure, the $\ell_0$-(pseudo)norm can be used;
however, for practical reasons, 
it is usually approximated by various techniques.
In this paper, the $\ell_1$-norm is used as the \qm{closest} convex surrogate of the $\ell_0$-norm.
%
Formally, the minimization problem can be written as
\begin{equation}
\argmin_{\c} \norm{\c}_1 \text{\ s.t.\ } \c \in \Gamma, 
\label{eq:problem_syn}
\end{equation}
where $\hspace{-0.02em}\Gamma \hspace{-0.2em}\subset\hspace{-0.2em} \CC^P\hspace{-0.2em}$ is a (convex) set of feasible solutions defined as
\begin{equation}
\Gamma = \{\c\ |\ \Mr \syn\c = \Mr\y, \Mh \syn\c \geq \tc, \Ml \syn\c \leq -\tc\},
\label{eq:Gamma}
\end{equation}
where $D : \CC^P \rightarrow\RR^N$ is the synthesis operator of a Parseval tight frame, with $P \geq N$
\cite{christensen2003,Kitic2015:Sparsity.cosparsity.declipping,ZaviskaRajmicPrusaVesely2018:RevisitingSSPADE}.
More specifically, throughout this paper, the Discrete Gabor Transform (DGT) is used in place of the time-frequency transform.
This transform is also commonly known as the Short-Time Fourier Transform (STFT).
The particular setting of the transform is described in Sec.\ \ref{sec:experiments}. 


To find an appropriate algorithm to solve \eqref{eq:problem_syn}, it is convenient to rewrite the problem to an unconstrained form: 
\begin{equation}
\argmin_{\c} \norm{\c}_1 + \iota_\Gamma(\c), 
\label{eq:unconst}
\end{equation}
where the hard constraint from \eqref{eq:problem_syn} is replaced by the indicator function of the set $\Gamma$ defined as
\begin{equation}
\iota_\Gamma (\c) = \left\{
	\begin{array}{ll}
			0 & \text{for } \c \in \Gamma, \\
			+\infty & \text{for } \c \notin \Gamma.
	\end{array}
\right.
\label{eq:indicator_func}
\end{equation}

Problem \eqref{eq:unconst} on its own does not support the incorporation of any additional (psychoacoustical) information. 
However, it is possible to replace the $\ell_1$-norm with the weighted $\ell_1$-norm by exploiting a~vector of non-negative weights $\w \in \RR^P$.
The minimization task will then attain the following form:
\begin{equation}
\argmin_{\c} \norm{\w\odot\c}_1 + \iota_\Gamma(\c).
\label{eq:weighted_unconst}
\end{equation}
Here the $\odot$ symbol represents the elementwise product.
The larger an element in $\w$ is, the more the respective coefficient in $\c$ gets pushed towards zero in the optimization.


%
 %
%

\section{Algorithmic solution}
\label{sec:alg_solution}
The Douglas-Rachford algorithm (DRA) \cite{combettes2007douglas} is able to find the minimizer of a~sum of two convex functions.
Problem \eqref{eq:weighted_unconst} is exactly an instance of such a~setup, since both the summands in the objective function are convex (although not strictly convex).
The algorithm is presented in Alg.\ \ref{alg:DR.declipping}.
\begin{algorithm}
	\DontPrintSemicolon
	\SetAlgoVlined
	\KwIn{Set starting point $\c^{(0)} \in \CC^P$, weights $\w\in\RR^P$.\\
				Set parameters $\lambda =1, \gamma > 0$.
	}
	\For{$i=0,1,\dots$\,}{
		$\tilde{\c}^{(i)} = \proj_{\Gamma}\c^{(i)}$ \;
		$\c^{(i+1)} = \c^{(i)} + \lambda \left(\soft_{\gamma\w}(2\tilde{\c}^{(i)} - \c^{(i)}) - \tilde{\c}^{(i)} \right)$ \;
	}
	\KwRet{$\c^{(i+1)}$}
	\caption{Douglas-Rachford algorithm solving \eqref{eq:weighted_unconst}}
	\label{alg:DR.declipping}
\end{algorithm}

The algorithm iterates over two main steps.
The first is the projection on $\Gamma$, which corresponds to the proximal operator of 
$\iota_\Gamma$, and the soft thresholding $\soft_{\gamma\w}$ with the vector of thresholds $\gamma\w$, which coincides with the proximal operator of the weighted $\ell_1$-norm.

Since only the Parseval tight frames are considered, the projection step can be computed via
the explicit formula
\cite{ZaviskaRajmicPrusaVesely2018:RevisitingSSPADE,RajmicZaviskaVeselyMokry2019:New_gerenalized_projection_declipping_preprint}:
\begin{equation}
\proj_\Gamma(\z) = \z - \syn^*\left(\syn\z-\proj(\syn\z)\right),
\label{eq:projection}
\end{equation}
%
where $D^* : \RR^N \rightarrow \CC^P$ is the analysis operator, 
and the inner projection step is a simple projection in the time domain, which is computed elementwise as:
\begin{equation}
\Big(\proj(\z)\Big)_n = \left\{
	\begin{array}{ll}
		y_n & \text{for } n \in R, \\
		\max(\tc, z_n) & \text{for } n \in H, \\
		\min(-\tc, z_n) & \text{for } n \in L.
	\end{array}
	\right.
\label{eq:proj_time_short}
\end{equation}



The soft thresholding, as the second main step of the DR algorithm, is computed according to
%
\begin{equation}
\soft_{\gamma\w}(\z) = \sgn(\z)\odot\max(|\z| - \gamma\w, 0).
\label{eq:soft_thr}
\end{equation}
%

%

\section{Choice of Weights}
\label{sec:weights}

The Douglas-Rachford algorithm for audio declipping expects the weighting vector $\w$ at the input.
The performance of the algorithm is greatly influenced by this vector. 
Effectively, weights affect the signal coefficients during the soft thresholding step; 
recalling \eqref{eq:soft_thr}, the higher weight belongs to a particular coefficient, the more shrunk the coefficient will become. 
Therefore, it is possible to encourage selecting the important coefficients 
and, on the other hand, discourage some others
(for example, discourage selecting higher frequency components introduced by clipping and thus not likely to be present in the original signal).

Three possible approaches to constructing the vector of weights $\w$ are proposed below.

\begin{figure}[t]%
\input{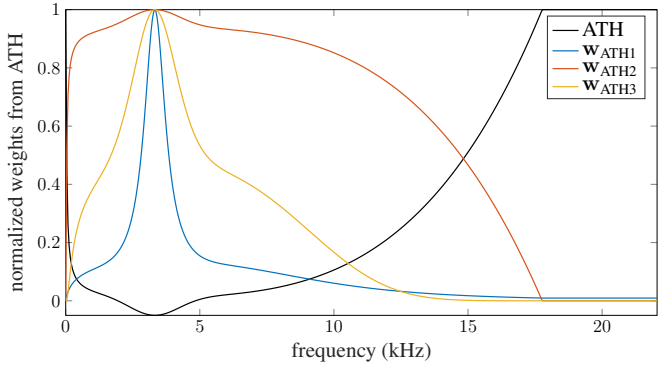}%
\vspace{-0.7em}\caption{Peak-normalized ATH curve and peak-normalized weights computed according to the three options in \eqref{eq:ath_weights}.\vspace{-1em}}%
\label{fig:weights_ath2}%
\end{figure}

\subsection{Absolute Threshold of Hearing}

The human audio perception ranges from 20\,Hz to ca 20\,kHz. 
It is also commonly known that the human ear
is more sensitive to frequencies around 2\,--\,5 kHz compared to the
higher and lower frequencies \cite{Psychoacoustics1999}.
This phenomenon was first characterized in 1933 by Fletcher and Munson as the Equal-loudness contours
and the most recent definition from 2003 is listed in the standard ISO226:2003 \cite{ISO_ATH}.
The equal-loudness contours indicate the frequency dependency of the sound pressure level of pure tone at a given frequency 
that is perceived by humans as loud as 1\,kHz pure tone associated with the same contour.
The minimal loudness at which a~harmonic sound is perceived is called the Absolute Threshold of Hearing (ATH) \cite{AudioSignalProcessingAndCoding}
and 
it can be approximated using the following equation \cite{Terhardt1979:CalculatingVirtualPitch}:
%
\begin{equation}
T_q(f) = 3.64\, g^{-0.8} - 6.5 \,\mathrm{e}^{-0.6 \left(g-3.3\right)^2} + 10^{-3} g^4, 
\label{eq:ATH_2}
\end{equation}
where $g = f/1000$, i.e.\ it is the frequency in kHz.

The main idea of using the ATH for weighting the coefficients in \eqref{eq:weighted_unconst} is to eliminate the negative effects of clipping,
especially at frequencies where the human hearing is most sensitive.
To do so, large weights should correspond to frequencies with the low respective ATH values and vice versa.
There is a~plethora of ways how to assign a weight to a corresponding frequency, based on the ATH curve.
We examine the following three options:
\begin{subequations}\label{eq:ath_weights}
	\begin{equation}
		\wa{1} = (\t - \min(\t) + 1)^{-1},
	\label{eq:ath_weighs1}
	\end{equation}
	\begin{equation}
		\wa{2} = -\t + \tau,
	\label{eq:ath_weights2}
	\end{equation}
	\begin{equation}
		\wa{3} = 2\cdot10^{-5}\cdot10^{(-\t + \tau)/{20}},
	\label{eq:ath_weights3}
	\end{equation}
\end{subequations}
where
$\t$ represents the vector of the ATH values for equispaced frequencies computed according to \eqref{eq:ATH_2},
and $\tau$ is the
parameter setting the maximum value of the ATH in dB.
Notice that $\wa{3}$ is basically $\wa{2}$ converted from dB(SPL) to the acoustic pressure in Pa.
All the operations in \eqref{eq:ath_weights} are conducted element-wise.

As the last step, the weights are always peak-normalized, so that the highest value of the weights is 1.
The normalization does not influence the result, but it affects the speed of convergence 
and also the proper setting of the parameter $\gamma$ in Alg.\ \ref{alg:DR.declipping}.

Normalized weights computed from the ATH according to \eqref{eq:ath_weights} with $\tau=100$, along with the original ATH contour, are shown in Fig.~\ref{fig:weights_ath2}.

\subsection{Global masking threshold}

\begin{figure}[t]%
\input{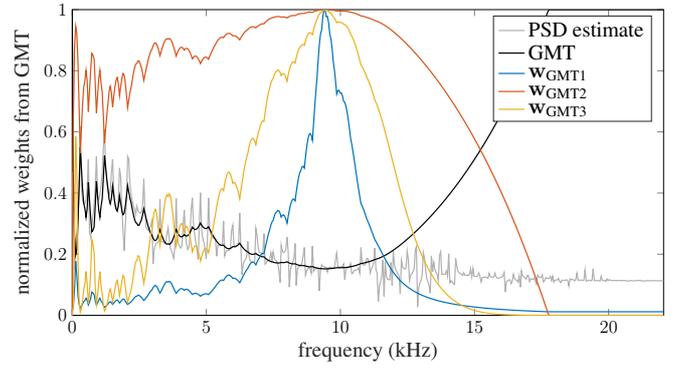}%
\vspace{-0.7em}\caption{Input DFT spectrum and corresponding Global masking threshold from which the weights are computed according to \eqref{eq:ath_weights}.\vspace{-1em}}%
\label{fig:weights_gmt}%
\end{figure}

Apart from the ATH, another phenomenon responsible for human auditory perception is \emph{simultaneous masking}, 
where the presence of a certain spectral component with high energy (the masker) masks another spectral component with lower energy (the maskee) \cite{Psychoacoustics1999}.
The masker makes the maskee inaudible to humans, although the component is physically present.

Combining this phenomenon with the ATH gives the \emph{global masking threshold} (GMT), which is a curve that indicates the minimum sound pressure level
that a spectral component has to possess in order not to be masked by other spectral components. 
In other words, spectral components below the GMT 
will be evaluated as imperceptible. 

The information contained in the GMT can be used to focus on restoring perceptually important components of the signal,
while tolerating less audible components because they are masked and thus not perceived.
Consequently, the weights should be constructed in a similar way to the case of ATH in the previous section, 
i.e.\ low values of GMT should produce large weights and vice versa.
The same three possibilities shown in \eqref{eq:ath_weights} were used, 
only the GMT plays the role of $\vect{t}$ now.
The resulting vectors of weights are $\w_\text{GMT1}$, $\w_\text{GMT2}$ and $\w_\text{GMT3}$.

The GMT itself is computed by a~slightly modified MPEG Psychoacoustic Model 1, where all the maskers are treated as the so-called tonal maskers \cite{Bosi2003:DigitalAudioCoding}.
In
brief, the GMT (for a~fixed time frame of the input signal) is obtained as follows:
First, the signal is weighted with the Hann window and the power spectral density (PSD) is estimated using the FFT. 
From the PSD the maskers are identified, and the respective individual masking thresholds
are calculated.
Finally, the GMT is obtained as a power-additive combination of the ATH and the just described individual thresholds.
An example of the GMT, the corresponding PSD estimate and the constructed weights are illustrated in Fig.~\ref{fig:weights_gmt}.

Recall that the ground-truth signal is not known in practice.
Computing the GMT from the observed (i.e.\ clipped) signal $\y$ may yield a biased estimate of the significant spectral components, 
especially for harsh clipping thresholds.
Therefore, a~way must be found to obtain a more correct GMT for use in \eqref{eq:weighted_unconst}.
We tackle this problem by first computing the declipping problem by simple $\ell_1$ minimization without weighting,
and we use the recovered signal as the basis for the GMT estimation and generation of the weights in \eqref{eq:weighted_unconst}. 

\subsection{Simple parabola}
The third option is based on the idea that most of the energy in audio signals is concentrated at lower frequencies
and that clipping introduces artificial higher harmonics that were not present in the original signal.
Consequently, we consider the option of constructing the weights in such a way that it will suppress the higher harmonics and preserve the low frequencies.

A~simple and effective approach to addressing this issue is to weight the coefficients with the identity curve. 
Better restoration results are obtained when a second-order polynomial is used instead of the identity.
Formally, these weights are obtained as
%
$\w_{\text{P}} = \vect{m} \odot \vect{m}$,
where $\vect{m} = [1, \dots, M]$, with $M$ denoting the number of frequency channels of the DGT. 

\section{Experiments and Results}
\label{sec:experiments}

\begin{figure}[t]%
%
%
\definecolor{mycolor1}{rgb}{0.00000,0.44700,0.74100}%
\definecolor{mycolor2}{rgb}{0.85000,0.32500,0.09800}%
\definecolor{mycolor3}{rgb}{0.92900,0.69400,0.12500}%
\definecolor{mycolor4}{rgb}{0.49400,0.18400,0.55600}%
\definecolor{mycolor5}{rgb}{0.46600,0.67400,0.18800}%
\begin{tikzpicture}[scale=0.57]

\begin{axis}[%
width=5.372in,
height=3.5in,
at={(0.462in,0.517in)},
scale only axis,
xmin=0.05,
xmax=0.95,
xlabel style={font=\Large\color{white!15!black}},
xlabel={$\text{clipping threshold }\theta{}_\text{c}$},
xticklabel style={font=\large},
ymin=0,
ymax=30,
yticklabel style={font=\large},
ylabel style={font=\Large\color{white!15!black}},
ylabel={$\Delta\text{SDR (dB)}$},
axis background/.style={fill=white},
xmajorgrids,
ymajorgrids,
legend style={at={(0.03,0.97)}, anchor=north west, legend cell align=left, align=left, draw=white!15!black, font=\large},
every axis plot/.append style={thick}
]
\addplot [color=mycolor1, mark=x, mark size=3pt, mark options={solid, mycolor1}]
  table[row sep=crcr]{%
0.1	1.71763731073241\\
0.2	4.54377078082827\\
0.3	7.50251675830948\\
0.4	9.82720142529285\\
0.5	11.4811378831795\\
0.6	13.2269630245376\\
0.7	14.5554383969816\\
0.8	14.7749172860894\\
0.9	15.4966504543065\\
};
\addlegendentry{noWeighting}

\addplot [color=mycolor2, mark=o, mark size=3pt, mark options={solid, mycolor2}]
  table[row sep=crcr]{%
0.1	3.97459661870579\\
0.2	7.82250998058606\\
0.3	9.93660299696574\\
0.4	11.609598601432\\
0.5	12.7437690584855\\
0.6	13.3788328980836\\
0.7	13.8872492377429\\
0.8	14.3306825077692\\
0.9	13.2681218842944\\
};
\addlegendentry{$\w_{\text{ATH1}}$}

\addplot [color=mycolor2, mark=square, mark size=3pt, mark options={solid, mycolor2}]
  table[row sep=crcr]{%
0.1	1.63306754806655\\
0.2	3.29280061597172\\
0.3	4.41089173909142\\
0.4	5.43201719597978\\
0.5	6.3312773432624\\
0.6	7.66051424166797\\
0.7	8.32953912608787\\
0.8	9.40778775943446\\
0.9	11.726328473478\\
};
\addlegendentry{$\w_{\text{ATH2}}$}

\addplot [color=mycolor2, mark=triangle, mark size=3pt, mark options={solid, mycolor2}]
  table[row sep=crcr]{%
0.1	3.54027503749491\\
0.2	5.29001754022508\\
0.3	5.86687411542661\\
0.4	6.06148430554634\\
0.5	5.96469276002987\\
0.6	5.65518102841538\\
0.7	5.42065595390298\\
0.8	5.9619010772259\\
0.9	5.82675798836221\\
};
\addlegendentry{$\w_{\text{ATH3}}$}

\addplot [color=mycolor3, mark=o, mark size=3pt, mark options={solid, mycolor3}]
  table[row sep=crcr]{%
0.1	4.47566411199997\\
0.2	9.08552953948744\\
0.3	12.1499192090171\\
0.4	14.4719904521796\\
0.5	16.2780335363193\\
0.6	17.7278130020427\\
0.7	18.698983751172\\
0.8	19.6629646427963\\
0.9	21.3619729844086\\
};
\addlegendentry{$\w_{\text{QMT1}}$}

\addplot [color=mycolor3, mark=square, mark size=3pt, mark options={solid, mycolor3}]
  table[row sep=crcr]{%
0.1	2.6408947179517\\
0.2	4.52746488539368\\
0.3	5.64776312665932\\
0.4	6.78714251907269\\
0.5	7.82887800025982\\
0.6	9.43963311107357\\
0.7	10.1463265675809\\
0.8	12.2317430264316\\
0.9	14.568282267508\\
};
\addlegendentry{$\w_{\text{QMT2}}$}

\addplot [color=mycolor3, mark=triangle, mark size=3pt, mark options={solid, mycolor3}]
  table[row sep=crcr]{%
0.1	7.72950754976377\\
0.2	9.65439362790497\\
0.3	10.3716568029486\\
0.4	10.7390273939295\\
0.5	11.3131500630479\\
0.6	11.8896889172813\\
0.7	12.425197932164\\
0.8	13.6218569556442\\
0.9	14.2193645821987\\
};
\addlegendentry{$\w_{\text{QMT3}}$}

\addplot [color=mycolor4, mark=x, mark size=3pt, mark options={solid, mycolor4}]
  table[row sep=crcr]{%
0.1	11.3680200512568\\
0.2	14.5009606344104\\
0.3	16.6452487342224\\
0.4	18.4975357723403\\
0.5	20.4312628217643\\
0.6	22.2097144658031\\
0.7	23.9539158084159\\
0.8	23.6559481487533\\
0.9	27.4835569344447\\
};
\addlegendentry{$\w_{\text{P}}$}

\addplot [color=mycolor5, mark=x, mark size=3pt, mark options={solid, mycolor5}]
  table[row sep=crcr]{%
0.1	12.8559212859794\\
0.2	15.8201634345788\\
0.3	17.0756550691254\\
0.4	19.2063473751034\\
0.5	20.1369383145744\\
0.6	20.893046725289\\
0.7	22.0141212603609\\
0.8	22.9751302622491\\
0.9	21.6074269942579\\
};
\addlegendentry{SPADE}

\end{axis}
\end{tikzpicture}%
\vspace{-0.7em}\caption{Average performance in terms of \dsdr{} for all proposed weighting variants of the $\ell_1$ minimization, plus the SPADE results for comparison.\vspace{-1em}}%
\label{fig:dSDR}%
\end{figure}
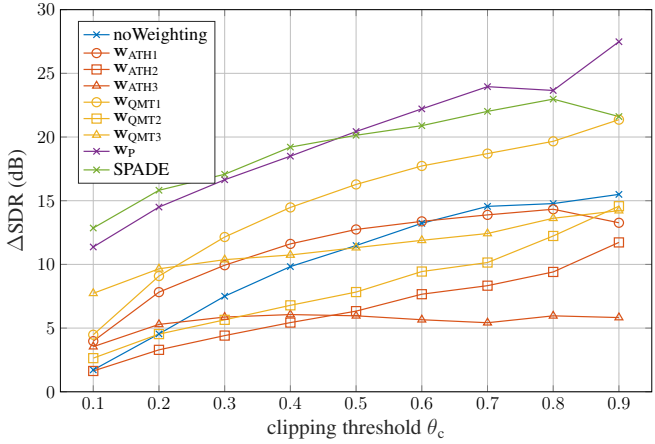

The experiments are designed to evaluate and compare the above-presented possibilities of incorporating psychoacoustical information into audio declipping.

A test set of 10 musical audio signals sampled at 44.1\,kHz with an approximate length of 7 seconds was
thoroughly selected from the EBU SQAM database%
\footnote{https://tech.ebu.ch/publications/sqamcd}
to be diverse in tonal character and sparsity with respect to the time-frequency transform used.
This way, the test set covers excerpts from the glockenspiel to the wind ensemble.
Each sound example was first peak-normalized and then clipped, using multiple clipping thresholds
$\tc = 0.1, \dots, 0.9$.  

Diverse sound excerpts were used with numerous clipping thresholds,
and our experience is that to obtain the best declipping results it would be necessary to tune the parameters of the DGT and the parameter $\gamma$ of the DRA
manually for each test instance.
For simplicity, these parameters were set to a~compromise among all the cases.
Specifically, Hann window 8192 samples long (i.e.\ 186\,ms) with 75\% overlap and 8192 frequency channels were used for the DGT.
Such a~setting generates a~Parseval frame.
The restoration algorithm was applied to a whole signal 
(i.e.\ the processing was not done frame-by-frame).
Inside the DRA,
the parameter $\gamma$ was set to 1 
and the number of iterations was set strictly to 1000.
Such a~number provided convergence in each of the test instances.

As a reference, the proposed approaches are compared with the SPADE algorithm \cite{Kitic2015:Sparsity.cosparsity.declipping}, 
which is, to the best of our knowledge, the state-of-the-art among the sparsity-based methods.
Unlike the proposed approach, SPADE is designed to process the signal frame-by-frame.
On purpose, our SPADE uses the same window shape and length and number of frequency channels.
Internal parameters of the SPADE were $r,s = 1$ and $\epsilon = 0.1$.

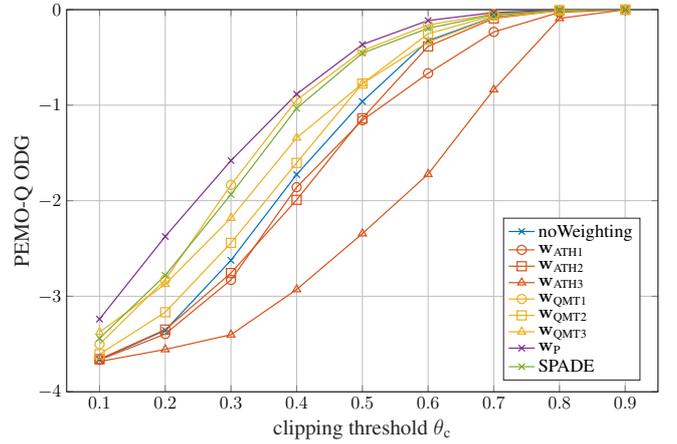
\begin{figure}[t]%
%
%
\definecolor{mycolor1}{rgb}{0.00000,0.44700,0.74100}%
\definecolor{mycolor2}{rgb}{0.85000,0.32500,0.09800}%
\definecolor{mycolor3}{rgb}{0.92900,0.69400,0.12500}%
\definecolor{mycolor4}{rgb}{0.49400,0.18400,0.55600}%
\definecolor{mycolor5}{rgb}{0.46600,0.67400,0.18800}%
\begin{tikzpicture}[scale=0.57]

\begin{axis}[%
width=5.372in,
height=3.5in,
at={(0.545in,0.517in)},
scale only axis,
xmin=0.05,
xmax=0.95,
xlabel style={font=\Large\color{white!15!black}},
xlabel={$\text{clipping threshold }\theta{}_\text{c}$},
xticklabel style={font=\large},
ymin=-4,
ymax=0,
ylabel style={font=\Large\color{white!15!black}},
ylabel={PEMO-Q ODG},
ylabel shift = -0.33em,
axis background/.style={fill=white},
ytick={-4, -3, -2, -1, 0},
yticklabel style={font=\large},
xmajorgrids,
ymajorgrids,
legend style={at={(0.97,0.03)}, anchor=south east, legend cell align=left, align=left, draw=white!15!black, font=\large},
every axis plot/.append style={thick}
]
\addplot [color=mycolor1, mark=x, mark size=3pt, mark options={solid, mycolor1}]
  table[row sep=crcr]{%
0.1	-3.66447405657278\\
0.2	-3.35717823104436\\
0.3	-2.62391420880684\\
0.4	-1.72572076641552\\
0.5	-0.961649829063214\\
0.6	-0.325464705956665\\
0.7	-0.076308814905363\\
0.8	-0.00693854232267164\\
0.9	-6.08750496123633e-05\\
};
\addlegendentry{noWeighting}

\addplot [color=mycolor2, mark=o, mark size=3pt, mark options={solid, mycolor2}]
  table[row sep=crcr]{%
0.1	-3.6668404290755\\
0.2	-3.39504511974618\\
0.3	-2.82914767226502\\
0.4	-1.85854575235413\\
0.5	-1.15810176354034\\
0.6	-0.66664646208063\\
0.7	-0.233494014169454\\
0.8	-0.0279477051789222\\
0.9	-0.000354155977226966\\
};
\addlegendentry{$\w_{\text{ATH1}}$}

\addplot [color=mycolor2, mark=square, mark size=3pt, mark options={solid, mycolor2}]
  table[row sep=crcr]{%
0.1	-3.65838086917323\\
0.2	-3.3496101045036\\
0.3	-2.75714927331682\\
0.4	-1.99078164061948\\
0.5	-1.13783522927538\\
0.6	-0.384545369394205\\
0.7	-0.091664595362073\\
0.8	-0.00990718866306892\\
0.9	-0.000108829605176552\\
};
\addlegendentry{$\w_{\text{ATH2}}$}

\addplot [color=mycolor2, mark=triangle, mark size=3pt, mark options={solid, mycolor2}]
  table[row sep=crcr]{%
0.1	-3.68287149552594\\
0.2	-3.55684298813903\\
0.3	-3.40264194096832\\
0.4	-2.9289066311727\\
0.5	-2.34446468537211\\
0.6	-1.72169506907246\\
0.7	-0.837086880980454\\
0.8	-0.0913525655058747\\
0.9	-0.000489344639575151\\
};
\addlegendentry{$\w_{\text{ATH3}}$}

\addplot [color=mycolor3, mark=o, mark size=3pt, mark options={solid, mycolor3}]
  table[row sep=crcr]{%
0.1	-3.50138190963536\\
0.2	-2.82665056892598\\
0.3	-1.83456995519395\\
0.4	-0.945780262336286\\
0.5	-0.430361754013131\\
0.6	-0.161217497791973\\
0.7	-0.0434872875760409\\
0.8	-0.00459661075679989\\
0.9	-5.37264619946853e-05\\
};
\addlegendentry{$\w_{\text{QMT1}}$}

\addplot [color=mycolor3, mark=square, mark size=3pt, mark options={solid, mycolor3}]
  table[row sep=crcr]{%
0.1	-3.60699625847989\\
0.2	-3.16743375910943\\
0.3	-2.44178457519623\\
0.4	-1.60397436252327\\
0.5	-0.773627058850768\\
0.6	-0.252773678518517\\
0.7	-0.0607276728554943\\
0.8	-0.00711727365460426\\
0.9	-7.31945822863622e-05\\
};
\addlegendentry{$\w_{\text{QMT2}}$}

\addplot [color=mycolor3, mark=triangle, mark size=3pt, mark options={solid, mycolor3}]
  table[row sep=crcr]{%
0.1	-3.37612250805497\\
0.2	-2.87334296853548\\
0.3	-2.1812720623502\\
0.4	-1.3413846742344\\
0.5	-0.774115922867073\\
0.6	-0.338953762799998\\
0.7	-0.0823232297812201\\
0.8	-0.0105526860235578\\
0.9	-9.22881815867527e-05\\
};
\addlegendentry{$\w_{\text{QMT3}}$}

\addplot [color=mycolor4, mark=x, mark size=3pt, mark options={solid, mycolor4}]
  table[row sep=crcr]{%
0.1	-3.23924090439372\\
0.2	-2.37416623259852\\
0.3	-1.57974030598994\\
0.4	-0.88371560255667\\
0.5	-0.364559803567069\\
0.6	-0.11420527433884\\
0.7	-0.0301036737924438\\
0.8	-0.00385291171007154\\
0.9	-2.45106372453563e-05\\
};
\addlegendentry{$\w_{\text{P}}$}

\addplot [color=mycolor5, mark=x, mark size=3pt, mark options={solid, mycolor5}]
  table[row sep=crcr]{%
0.1	-3.44036485092183\\
0.2	-2.78138810642455\\
0.3	-1.93408801505154\\
0.4	-1.03266406722922\\
0.5	-0.455097332921777\\
0.6	-0.195312174928171\\
0.7	-0.0513935886992801\\
0.8	-0.00630484404569387\\
0.9	-0.000146315493181248\\
};
\addlegendentry{SPADE}

\end{axis}
\end{tikzpicture}%
\vspace{-0.7em}\caption{Average PEMO-Q ODG values for all proposed weighting variants of the $\ell_1$ minimization algorithm, and the SPADE algorithm for comparison.\vspace{-1em}}%
\label{fig:pemoq}%
\end{figure}

\subsubsection{Evaluation using signal-to-distortion-ratio (SDR)}
The physical quality of restoration, i.e.\ how much the recovered signal $\hat{\x}$ is similar to the ground truth $\x$,
is evaluated by the $\dsdr$,
which is computed as a difference between the SDR values of the restored and the clipped signal, 
$\dsdr = \text{SDR}(\x, \hat{\x}) - \text{SDR}(\x, \y)$, 
and thus it expresses the SDR
improvement in dB.
%
The SDR for two signals $\u$ and $\vv$ is computed as
\begin{equation}
\text{SDR}(\u, \vv) = 10\log_{10}\frac{\norm{\u}^2_2}{\norm{\u-\vv}^2_2}.
\label{eq:sdr}
\end{equation}

Fig.~\ref{fig:dSDR} illustrates the average \dsdr{} values for all the proposed choices of weights (see Sec.~\ref{sec:weights}), 
in comparison with the pure non-weighted version of Algorithm \ref{alg:DR.declipping} and with the SPADE algorithm.
The resulting \dsdr{} values indicate that weighting with the GMT is a better approach than just a simple ATH curve in all the cases.
Also, the best variant of converting the GMT or ATH curves into the actual weight vector $\w$ seems to be the one using the inversion, Eq.\,\eqref{eq:ath_weighs1}.
Nevertheless, among all choices, the best results (in terms of \dsdr{}) are obtained by weights derived from a parabola. 

Comparing the $\ell_1$ model with the SPADE, the weights based on the parabola perform slightly better for mild clipping thresholds ($\tc\geq0.5$) 
while SPADE performs slightly better for harsher clipping thresholds ($\tc < 0.5$).

\subsubsection{Evaluation using PEMO-Q}

Since the topic of this paper is psychoacoustically motivated, the goal is to recover a signal
that would sound the best to the listener;
this requirement is not always quite in correspondence with the similarity in terms of the SDR.
In other words, the \dsdr{} is not the right objective measure and thus we use the \mbox{PEMO-Q} \cite{Huber:2006a} evaluator, 
which should be more consistent with the human auditory system.
The PEMO-Q output scale---ODG (the objective difference grade)---ranging from 0 to $-4$ is interpreted as:
\vspace{-0.2em}
\begin{center}
\small
\begin{tabular}{rl}
	 $0.0$ & Imperceptible \\
	$-1.0$ & Perceptible, but not annoying \\
	$-2.0$ & Slightly annoying \\
	$-3.0$ & Annoying \\
	$-4.0$ & Very annoying.
\end{tabular}
\end{center}
\vspace{-0.2em}

The PEMO-Q ODG values for the experiment are shown in Fig.~\ref{fig:pemoq}.
A comparison of Figs.~\ref{fig:dSDR} and \ref{fig:pemoq} confirms the assertion that \dsdr{} does not correspond quite well to human perception (as modeled by the PEMO-Q).
The PEMO-Q results indicate that weights based on the ATH do not help to enhance the quality of restoration compared with the plain $\ell_1$ minimization.
Weighting based on the GMT, on the contrary, improves the ODG values of restored signals, 
with the option \eqref{eq:ath_weighs1} being the best out of the three and
even outperforming the SPADE algorithm for $\tc \geq 0.3$.
The best restoration quality, according to PEMO-Q ODG, is delivered by the $\ell_1$ minimization weighted with parabola.

The source codes are available at \url{http://www.utko.feec.vutbr.cz/~rajmic/software/declip_psychoacoust.zip}.



\section{Conclusion}
In this paper, a novel sparsity-based approach to audio declipping has been proposed
that incorporates psychoacoustic information.
The results indicate that weighting helps to improve the obtained quality of restoration not only in SDR but more importantly in the PEMO-Q ODG scale, 
which better corresponds to human audio perception. 
The best results were obtained by weighting with the inverse GMT curve and parabola,
which even outperform the SPADE algorithm in terms of the PEMO-Q ODG, while being ca $2.5\times$ faster. 

As a subject of future research, the combination of the GMT with parabola could be promising since both different options produce good results. 
We also plan to involve the psychoacoustic weighting into social sparsity based algorithms
and into the SPADE algorithm.






%

{
\bibliographystyle{IEEEtran}
\inputencoding{cp1250}
\bibliography{IEEEabrv,literatura}

\newcommand{\noopsort}[1]{} \newcommand{\printfirst}[2]{#1}
  \newcommand{\singleletter}[1]{#1} \newcommand{\switchargs}[2]{#2#1}
\begin{thebibliography}{10}
\providecommand{\url}[1]{#1}
\csname url@samestyle\endcsname
\providecommand{\newblock}{\relax}
\providecommand{\bibinfo}[2]{#2}
\providecommand{\BIBentrySTDinterwordspacing}{\spaceskip=0pt\relax}
\providecommand{\BIBentryALTinterwordstretchfactor}{4}
\providecommand{\BIBentryALTinterwordspacing}{\spaceskip=\fontdimen2\font plus
\BIBentryALTinterwordstretchfactor\fontdimen3\font minus
  \fontdimen4\font\relax}
\providecommand{\BIBforeignlanguage}[2]{{%
\expandafter\ifx\csname l@#1\endcsname\relax
\typeout{** WARNING: IEEEtran.bst: No hyphenation pattern has been}%
\typeout{** loaded for the language `#1'. Using the pattern for}%
\typeout{** the default language instead.}%
\else
\language=\csname l@#1\endcsname
\fi
#2}}
\providecommand{\BIBdecl}{\relax}
\BIBdecl

\bibitem{Tan2003}
\BIBentryALTinterwordspacing
C.-T. Tan, B.~C.~J. Moore, and N.~Zacharov, ``The effect of nonlinear
  distortion on the perceived quality of music and speech signals,''
	\emph{J.~Audio Eng. Soc}, vol.~51, no.~11, pp. 1012--1031, 2003.
\BIBentrySTDinterwordspacing

\bibitem{Malek2013:Blind.compensation}
J.~M\'{a}lek, ``Blind compensation of memoryless nonlinear distortions in
  sparse signals,'' in \emph{21st EUSIPCO}, Sept 2013, pp. 1--5.

\bibitem{Tachioka2014:Speech.recog.performance}
Y.~Tachioka, T.~Narita, and J.~Ishii, ``Speech recognition performance
  estimation for clipped speech based on objective measures,'' \emph{Acoustical
  Science and Technology}, vol.~35, no.~6, pp. 324--326, 2014.

\bibitem{Patent_clipping_amplifier}
T.~Ikoma, ``Apparatus for avoiding clipping of amplifier,'' U.S. Patent
  US4\,581\,589A, 1984.

\bibitem{javevr86}
A.~J. E.~M. Janssen, R.~N.~J. {V}eldhuis, and L.~B. {V}ries, ``{A}daptive
  interpolation of discrete-time signals that can be modeled as autoregressive
  processes,'' \emph{{IEEE} {T}rans. {A}cous., {S}peech and {S}ignal
  {P}roc.}, vol.~34, 1986.

\bibitem{abel91_declipping}
J.~Abel and J.~Smith, ``Restoring a clipped signal,'' in \emph{Acoustics,
  Speech, and Signal Processing, 1991. ICASSP}, 1991, pp.\ 1745--1748, vol.\,3.

\bibitem{Godsill2001:StatisticalAudioRestoration}
\BIBentryALTinterwordspacing
S.~J. Godsill, P.~J. Wolfe, and W.~N. Fong, ``Statistical model-based
  approaches to audio restoration and analysis,'' \emph{Journal of New Music
  Research}, vol.~30, no.~4, pp. 323--338, 2001. 
\BIBentrySTDinterwordspacing

\bibitem{FongGodsill2001:MonteCarlo}
W.~Fong and S.~Godsill, ``Monte carlo smoothing for non-linearly distorted
  signals,'' in \emph{Acoustics, Speech and Signal Processing (ICASSP), 2019 IEEE International Conference on}, 
	2001, pp. 3997--4000.

\bibitem{Chantas2018:Inpainting_Variational_Bayesian_Inference}
G.~{Chantas} et al. 
	``Sparse audio inpainting with variational bayesian inference,'' in \emph{2018 IEEE
  Intl.\ Conf.\ on Consumer Electronics}, 2018.

\bibitem{Takahashi2013:Hankel_matrix_declipping}
T.~{Takahashi}, K.~{Konishi}, and T.~{Furukawa}, ``Hankel structured matrix
  rank minimization approach to signal declipping,'' in \emph{21st European
  Signal Processing Conference (EUSIPCO 2013)}, Sep. 2013, pp. 1--5.

\bibitem{Adler2011:Declipping}
A.~Adler et al. 
	``A~constrained matching pursuit approach to audio declipping,'' in
  \emph{Acoustics, Speech and Signal Processing (ICASSP), 2011 IEEE
  International Conference on}, 2011, pp. 329--332.

\bibitem{Weinstein2011:DeclippingSparseland}
\BIBentryALTinterwordspacing
A.~J. Weinstein and M.~B. Wakin, ``Recovering a clipped signal in sparseland,''
  \emph{CoRR}, vol. abs/1110.5063, 2011.
\BIBentrySTDinterwordspacing

\bibitem{Defraene2013:Declipping.perceptual.compressed.sensing}
B.~Defraene et al., 
	``Declipping of audio signals using perceptual compressed sensing,'' 
	\emph{IEEE Transactions on Audio, Speech, and Language Processing}, 
	vol.~21, no.~12, pp. 2627--2637, Dec 2013.

\bibitem{SiedenburgKowalskiDoerfler2014:Audio.declip.social.sparsity}
K.~Siedenburg, M.~Kowalski, and M.~Dorfler, ``Audio declipping with social
  sparsity,'' in \emph{Acoustics, Speech and Signal Processing (ICASSP), 2014
  IEEE International Conference on}.\hskip 1em plus 0.5em minus 0.4em\relax
  IEEE, 2014, pp. 1577--1581.

\bibitem{Kitic2013:Consistent.iter.hard.thresholding}
S.~Kiti{\'c} et al., 
	``Consistent iterative hard thresholding for signal
  declipping,'' in \emph{Acoustics, Speech and Signal Processing (ICASSP), 2013
  IEEE International Conference on}, May 2013, pp. 5939--5943.

\bibitem{KiticBertinGribonval2014:AudioDeclippingCosparseHardThresholding}
S.~Kiti{\'c}, N.~Bertin, and R.~Gribonval, ``Audio declipping by cosparse hard
  thresholding,'' in \emph{2nd Traveling Workshop on Interactions between
  Sparse models and Technology}, 2014.

\bibitem{Kitic2015:Sparsity.cosparsity.declipping}
------, ``Sparsity and cosparsity for audio declipping: a flexible non-convex
  approach,'' in
  \emph{Latent Variable Analysis and Signal Separation}, Liberec, Czech Republic,
  Aug. 2015.

\bibitem{ZaviskaRajmicPrusaVesely2018:RevisitingSSPADE}
P.~Z{\'a}vi{\v{s}}ka, P.~Rajmic, Z.~Pr{\r{u}}{\v{s}}a, and V.~Vesel{\'y},
  ``Revisiting synthesis model in sparse audio declipper,'' in \emph{Latent
  Variable Analysis and Signal Separation}.
  Guildford, UK, 2018, pp. 429--445.

\bibitem{ZaviskaRajmicMokryPrusa2019:SSPADE_ICASSP}
\BIBentryALTinterwordspacing
P.~{Z{\'a}vi{\v s}ka}, O.~{Mokr{\'y}}, and P.~{Rajmic}, 
``A~proper version of synthesis-based sparse audio declipper,'' 
in \emph{Acoustics, Speech and Signal Processing (ICASSP), 2019 IEEE Intl.\ Conf.\ on}, 2019, pp.\ 591--595.

\bibitem{BilenOzerovPerez2015:declipping.via.NMF}
C.~Bilen, A.~Ozerov, and P.~Pérez, ``Audio declipping via nonnegative matrix
  factorization,'' in \emph{Applications of Signal Processing to Audio and
  Acoustics (WASPAA), 2015 IEEE Workshop on}, Oct 2015, pp. 1--5.

\bibitem{Bilen2018:NTF_audio_inverse_problems}
C.~{Bilen}, A.~{Ozerov}, and P.~{Pérez}, ``Solving time-domain audio inverse
  problems using nonnegative tensor factorization,'' \emph{IEEE Transactions on
  Signal Processing}, vol.~66, no.~21, pp.\ 5604--5617, Nov 2018.
	
\bibitem{HonigWerman2018:Image.decliping}
S.~{Honig} and M.~{Werman}, ``Image declipping with deep networks,'' in
  \emph{2018 25th IEEE Intl.\ Conf.\ on Image Processing (ICIP)},
  Oct.\ 2018.

\bibitem{combettes2007douglas}
P.~Combettes and J.~Pesquet, ``{A Douglas--Rachford splitting approach to
  nonsmooth convex variational signal recovery},'' \emph{IEEE Journal of
  Selected Topics in Signal Processing}, vol.~1, no.~4, pp. 564--574, 2007.

\bibitem{christensen2003}
O.~Christensen, \emph{An Introduction to Frames nad Riesz Bases}.\hskip 1em
  plus 0.5em minus 0.4em\relax Boston-Basel-Berlin: Birkhäuser, 2003.

\bibitem{RajmicZaviskaVeselyMokry2019:New_gerenalized_projection_declipping_preprint}
\BIBentryALTinterwordspacing
P.~Rajmic, P.~Záviška, V.~Veselý, and O.~Mokrý, ``New generalized projection
  speeds up audio declipping,'' Mar. 2019, preprints 2019. 
\BIBentrySTDinterwordspacing

\bibitem{Psychoacoustics1999}
H.~Zwicker, E;~Fastl, \emph{Psychoacoustics: Facts and Models}, 2nd~ed.\hskip
  1em plus 0.5em minus 0.4em\relax New York: Springer, 1999.

\bibitem{ISO_ATH}
\emph{Acoustics -- Normal equal-loudness-level contours}, ISO Std. 226:2003.

\bibitem{AudioSignalProcessingAndCoding}
A.~Spanias, T.~Painter, and V.~Atti, \emph{\BIBforeignlanguage{English
  (US)}{Audio Signal Processing and Coding}}.\hskip 1em plus 0.5em minus
  0.4em\relax John Wiley \& Sons, Inc., 12 2005.

\bibitem{Terhardt1979:CalculatingVirtualPitch}
\BIBentryALTinterwordspacing
E.~Terhardt, ``Calculating virtual pitch,'' \emph{Hearing Research}, vol.~1,
  no.~2, pp. 155 -- 182, 1979.
\BIBentrySTDinterwordspacing

\bibitem{Bosi2003:DigitalAudioCoding}
R.~Bosi, M.;~Goldberg, \emph{Introduction to Digital Audio Coding and
  Standards}.\hskip 1em plus 0.5em minus 0.4em\relax Kluwer Academic
  Publishers, 2003.

\bibitem{Huber:2006a}
R.~Huber and B.~Kollmeier, ``{PEMO-Q---A} new method for objective audio
  quality assessment using a model of auditory perception,'' \emph{IEEE Trans.
  Audio Speech Language Proc.}, vol.~14, no.~6, Nov.\ 2006.

\end{thebibliography}
}


\end{document}